# Multi-Authority Ciphertext-Policy Attribute Based Encryption With Accountability

Wei Zhang[1], Yi Wu[1] ,Zhishuog Zhang[1],Hu Xiong[1] and Zhiguang Qin[1]

[1] University of Electronic Science and Technology of China
Chengdu, 610054, China
[e-mail: zhangving@uestc.edu.cn]
*Corresponding author: Hu Xiong

**Abstract.** Attribute-based encryption (ABE) is a promising tool for implementing fine-grained access control.To solve the matters of security in single authority, access policy public, not traceable of malicious user,we proposed a scheme of multi-authority. Moreover, multi-authority may bring about the collusion of different authorities.In order to solve these problem,we proposed a scheme of access tree structure with policy hidden and access complex.Once the private key is leaked, our scheme can extract the user ID and find it.If the authorities share their information with each other,the scheme avoid them to combine together to compute the key information and decrypt the ciphertext.Finally,the scheme proved to be secure under selective-set of IND-CPA.

**Keywords:** multi-authority · collusion · traceable · access tree

## 1 Introduction

With the rapid development of cloud storage technology, more and more individuals and enterprises tend to store the private data in the cloud server. To ensure the security of the information is facing more severe challenges. Sahai and Waters[1] first proposed the notion of attribute-based encryption(ABE) in 2005,in which a user is described with a set of attributes and the secret key is generated by the authority on the basis of these attributes. The user decrypts the ciphertext base on the threshold access structure in the above system. In order to achieve more complex access structures, Goyal et al.[2] proposed the scheme of  key policy attribute-based encryption (KP-ABE) in which the access policy is embedded in the key.Except for KP-ABE, Bethen court et al.[3]proposed the notion of ciphertext-policy attribute-based encryption (CP-ABE).Different from KP-ABE, the access policy is embedded in the ciphertext related to attributes.In both of the above ABE notions, a user can decrypt the information only if the access policy satisfies the attributes in key or ciphertext.

## 1.1 Multi-authority attribute-based encryption

In traditional single attribute authority ABE,since the user or authority is not fully trusted in the system，once the authority is malicious and brought through by an attacker，it will bring a huge threaten to all the users in the system.Moreover.single attribute authority to assign the attributes may hit a bottleneck on efficiency more easily since the algorithm and data grows. To solve the problem,in 2007 Chase et al.[4] proposed an Optimization Schemes of ABE with multiple authorities. Different attribute authority charges different attributes respectively and issues attribute keys to users,furthermore reduces the risk of trust assumption in single authority.Despite all this,there is still a central attribute authority which is fully trusted and manages the information specifically defines the identities of users and the state of each attribute authority.

　　As a result of multiple authorities, Another derive challenge is how to prevent collusion attacks by users or the multiple authorities.Different users with different access structure may combine together to generate a new decryption key towards another access structure.Different attribute authorities may also share its own information with each other to gain the knowledge of the decryption key.

## 1.2 Attribute-based encryption with accountability

In traditional ABE，since the user or authority is not fully trusted in the system，once the private key or other secret information leaked, it is unknown who should be responsible for it .To address this problem, Li et al.[5] first presented a traceable multi-authority CP-ABE, but it is confined to expressing a strict "AND" gates with wildcard policy. However, Liu et al.[6] pointed out that the malicious user cannot be traced simply by the approach in [7]. Later, Zhou et al.[8] proposed a traceable multi-authority CP-ABE scheme in which the policies can be expressed in any monotone access structures.

## 2 Definitions

### 2.1 Bilinear maps

we define multiplicative cyclic groups $G_1, G_2$ with prime order and set bilinear map $e:G_1 \times G_1 \rightarrow G_2$.g is the generator of $G_1$. If $e$ has the following properties:

　　1.Bilinear: In $e:G_1 \times G_1 \rightarrow G_2$, $e(aP,bQ) = e(P,Q)^{ab}$ for all $P,Q \in G_1$ and all $a,b \in Z_p$.

　　2.Non-degenerate: The map does not send all pairs in $G_1 \times G_1$ to the identity in $G_2$.Observe that since $G_1, G_2$ are groups of prime order this implies that if $P$ is a generator of $G_1$,then $e(P,P)$ is a generator of $G_2$.

　　3.Computable: There exists an efficient algorithm to compute $e(P,Q)$ for for all $P,Q \in G_1$.

## 2.2 Access tree structure

In CP-ABE with access tree structure, a set of attributes is embedded in the ciphertext as a structure of a tree.Each leaf node of the tree stores the attribute and the value assigns to it. Moreover,parent node encrypts its secret value and transmits to leaf node,only the users with these attributes have the ability to decrypt it.Each non-leaf node maintains a threshold gate *n/m,m* is described by the number of its children and n is the threshold value, representing the minimum number of attributes required to decrypt it. As shown in Figure 1, the threshold of Node 2 is 3/3 means that only the user who has the attributes "*male*"、"*computer*" and "*Grade 2*" can decrypt this node.The threshold of Node 4 is 1/2 means that the user who has the attributes "*network*" or "*cloud*" can solve the problem.

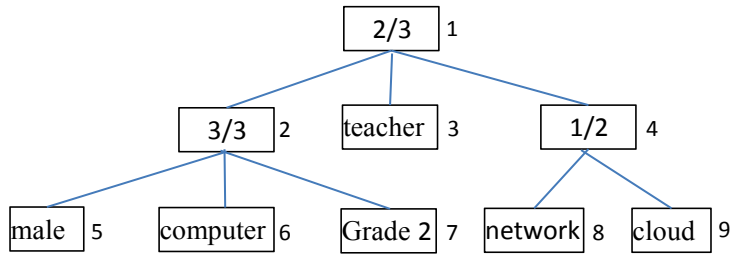

Figure 1 Tree structure

Each node has a polynomial $p_x$ to describe it. Assume that the degree of $p_x$ is $d_x$,which is one less than the threshold value n,that is $d_x=n-1$.The secret value of this node is $p_x(0)$ and for its children from left to right,the secret value is $p_x(1)$、$p_x(2)$... etc.So for for any other node(except for root node), $p_x(0)=p_{parent(x)}(index(x))$.

## 2.3 Syntax and security model

A multi-authority ciphertext-policy attribute based encryption with accountability scheme should consist of following five algorithms.

**1.INIT:**Select a security parameter $\kappa$ and run the algorithm to output a master public key(*mpk*) and a master secret key(*msk*).

**2.Key Gen:**An interactive between authority and user.Seting $P(id,\tau)$ is the public input where id is the user's global identify in the system and $\tau$ is the access tree associate to the user.The private input is the *msk*.At the end of the interactive,a private key $d_{id,\tau}$ of $P(id,\tau)$ is extracted.

**3.Encryption:**Taking the *mpk*, an attributes set $\omega$ and a message *M* as parameters and works out the ciphertext *C*.

**4.Decryption:**Taking $d_{id,\tau}$ of $P(id,\tau)$,ciphertext *C* as parameters and output the plaintexts *M* if $\omega \in \tau$.

**5.Trace:**Taking a well-formed private key $d_{id,\tau}$ and outputs the associated id.

The above proposed scheme should be indistinguishable under chosen plaintext Attack(*IND-CPA*).In this situation, excepting for encryption algorithm is leaked,the selected plaintexts and associated ciphertexts are also known by an attacker.In this scheme we extend the *Selective-set (SS)* model to our setting. The basic mentality to security proof of *IND-CPA* is as follows.

**1.INIT:** The adversary declares an attributes set $\omega^*$, which is to be challenged.

**2.Setup:** The challenger runs the algorithm and outputs public parameters to the adversary.

**3.Phase 1:** The adversary runs the algorithm with the inputs of pairs of *($id_j, \tau_j$)*, in which $\omega^*$ is not the subset of $\tau_j$ for all *j*.

**4. Challenge:** The adversary submits two messages $M_0, M_1$ with equal length. The challenger flips a coin as *v,v=0* or *v=1*, encrypts $M_v$ with $\omega^*$, output the ciphertext and passes to the adversary.

**5.Phase 2:** Repeat Phase 1.

**6.Guess:** The adversary outputs a guess $v'$ of *v*.

The advantage of adversary in this game is defined as $Pr[v'=v]-1/2$.

## 3.Our scheme

In this multi-authority system, set $A_1$、$A_2...A_k$ as the *K* attribute authorities, which are in charge of the disjoint *N* attributes associate to users. Set CA as the single center authority, which is fully trusted and manages the information specifically defines the identities of users and the state of each attribute authority. CA maintains a secret function *V(x)* mapping id to $Z_p^*$ a table T, storing the user *ID* and *V(id)*.

**1.INIT:** We set $\{t_{k,i}\}$ as the *i-th* attribute in $A_k$ which is mapped to $Z_p^*$ where *k=1...K* and *i=1...N*. Choosing $se_1, se_2......se_k$ as the *PRF* seeds of each authority. Each authority *k* chooses a different $y_{k,u}$ for each user. Moreover, randomly choose $x,y,y_0,y_1,s_0,w,w_1,w_2,\theta \in Z_p^*$ and $Z,h,X,Y,Y_1,g_1,g_2,g_3,g_4 \in G_1$, set $X=g^x$, $Y=g^y$, $Y_1=g_3^{y_0}$, $g_1=g^{y_0}$, $E_1=e(g,g)^{y_1}$, $g_2=g^{w_1}$, $g_3=g^{w_2}$, $g_4=g_2g_3=g^w$, $Y_0=e(g,g)^{y_0}$.

Now set the master secret key as $msk=(x,y,y_0,y_1,s_0,w,w_1,w_2,\theta)$, and the master public key as $mpk=(E_1,Z,h,X,Y,Y_1,Y_0,g_1,g_2,g_3,g_4)$.

**2.Key Gen:** To Authority *k*, set $se_k$ and $\{t_{k,i}\}$ as the secret key and $T_{k,i}=g^{t_{k,i}}$ as public key where *i=1...N*. For user u, set $y_{k,u}=F_{sk}(u)$ and randomly choose polynomial *p* of *d-1* degree where *d* is the threshold of attributes to decrypt authority *k*. Set $p(0)=y_{k,u}$ and secret key $D_{k,i}=\{g_2^{p(i)/t_{k,i}}\}, i \in A_u$.

To user P with access tree structure $\tau_k$, Set $R = h^{s_0}X^\theta$ and provides it to CA with an interactive witness indistinguishable proof of knowledge of the *($s_0,\theta$)*. CA randomly chooses $s_1=V(id), r' \in Z_p^*$ and returns to P with $d'_{id,\tau_k}=(d'_1,d'_2,d'_3,d'_4,d'_5)$

where $d'_1 = (Y_1Rh^{s_1})^{\frac{1}{x}} \cdot (g^{id}Z)^{r'}$, $d'_2=X^{r'}, d'_3=s_1, d'_4=D_{k,i}$

To CA, set $se_k, y_{k,u}$ of all attribute authorities as private values and set $D_{ca}=g_2^{(y_0 - \Sigma_{i=1}^{K} y_{k,u})}$ as $d_5'$. P then chooses $r'' \in Z_p^*$ and compute

$d_{id,\tau}=(d_1,d_2,d_3,d_4,d_5)$

$=((d_1'/g^\theta \cdot (g^{id}Z)^{r''}), d_2' \cdot X^{r''}, d_3' + s_0, d_4', d_5')$

$=(Y_1 R h^{s_1 + s_0})^{\frac{1}{x}} \cdot (g^{id}Z)^r, X^r, s_1 + s_0, D_{k,i}, g_2^{(y_0 - \Sigma_{i=1}^{K} y_{k,u})})$ where $r=r' + r''$

and $i \in A_u$.

**3. Encryption:** For attribute set $A_c$, Randomly choose $s \in Z_p^*$ and set

$E'=g^s, \{E_{k,i} = T_{k,i}^s\}_{i \in A_C^k}, E=Me(g,g_4)^{sy_0}$. So the ciphertext is as

$C=(A_c, C_1, C_2, C_3, C_4, C_5)=(A_c, X^s, E', Z^s, E, E_{k,i})$.

**4. Decryption:** For each leaf node, we compute algorithm $e(E_{k,i}, D_{k,i})=e(g,g_2)^{p(i)s}$ if number of $A_c^k \cap A_u^k$ is more than threshold value. we can use Lagrange interpolation formula to obtain $e(g,g_2)^{p(0)s}$ for parent node x of each leaf node. Finally we can decrypt the tree with $e(g,g_2)^{p(0)s}= e(g,g_2)^{y_{k,u}s}=Y_{k,u}^s$. Since

$e(d_1,C_1)/(e(C_2,h)^{d_3} \cdot e(C_2^{id} \cdot C_3, d_2))=e(g^s,Y_1)=e(g^{sy_0},g_3)$

and $e(C_2, d_5) \cdot \prod_{k=1}^{K} Y_{k,u}^s = e(g^s, g_2^{(y_0 - \Sigma_{i=1}^{K} y_{k,u})}) \cdot \prod_{k=1}^{K} Y_{k,u}^s = e(g^s, g_2^{y_0}) = e(g^{sy_0}, g_2)$,

So we have

$e(C_2, d_5) \cdot (\prod_{k=1}^{K} Y_{k,u}^s) \cdot e(d_1,C_1)/(e(C_2,h)^{d_3} \cdot e(C_2^{id} \cdot C_3, d_2))= C_4/M$

**5. Trace:** If $d_{id,\tau}$ is leaked and well-formed, the key component of $d_3$ has the information of $s_1=V(id)$. Searching table $T$ in CA and output the corresponding id.

## 4. Secure proofs

This section will give the security proof for above ABE scheme of *IND-SS-CPA* under the *DBDH* assumption. First we give the definition of *DBDH* assumption.

For the given bilinear map $e:G_1 \times G_1 \to G_2$ and groups $G_1, G_2$ with prime order, letting g be a generator of $G_1$, DBDH assumption holds if two distributions $(g^a, g^b, g^c, e(g,g)^{abc})$ and $(g^a, g^b, g^c, e(g,g)^z)$ are indistinguishable in any polynomial-time adversary $\beta$, where $a,b,c,z$ are randomly chosen in $Z_p^*$. we flip a fair binary coin $\mu$ outside $\beta'$ view, if $\mu= 0$, the challenger sets the distribution as $(g^a, g^b, g^c, e(g,g)^{abc})$, otherwise as $(g^a, g^b, g^c, e(g,g)^z)$.

In other words, we say that if $\beta$ has advantage $\varepsilon$ against our scheme, then we can build a simulator that can solve *DBDH* problem with advantage $\varepsilon/2$.

**1. Init:** The simulator receives the target attributes set $\omega^*$ from adversary $\beta$.

**2.Set up:** Set $e(A,B)=e(g_1,g_4)$ (this implicitly set $wy_0 = ab$). $X=C=g^c, Z=C^w$.

**3.Phase 1:** Suppose $\beta$ requests a private key $(id, \tau)$ where $\tau(\omega^*)=0$, which means $\omega^* \notin \tau$. The simulator will receive $R=h^{s_0} \cdot X^\theta$ with an interactive witness indistinguishable proof of knowledge of the $(s_0, \theta)$ from $\beta$. It then selects $s_1, r_2, r_3 \in Z_p^*$. The simulator extract the private key as follows. We set the query private key as

$d_{queryid,\tau}=(d_{query1}, d_{query2}, d_{query3}, d_{query4}, d_{query5})$ in which

$d'_{query1}=(g^{id}Z)^{r_2} \cdot (Y_1Rh^{s_1})^{-w/id}, d'_{query2}=X^{r_2} \cdot (Y_1Rh^{s_1})^{-/id}$,

$d'_{query3}=s_1$,

For the attribute authorities with $A_u^k \cap A_c^k \geq d$, we set $p(0)=F_{sk}(u)=z_{k,u}b$, $\rho(0)=z_{k,u}$ where $z_{k,u}$ and polynomial $\rho$ are randomly chosen.

$d'_{query4}=g_4^{b\rho(i)/\beta_{k,i}}$ ($i \in A_u^k \cap A_c^k$), otherwise $g_4^{\rho(i)/\beta_{k,i}}$ ($i \in A_u^k - A_c^k$).

$d'_{query5}=g^{(\Sigma z_{k',u} - \Sigma F_{sk}(u))}$ where k' are honest authorities and k are corrupted authorities.

Let $r_2=r_1 + \log_g(Y_1Rh^{s_1})/(x \cdot id)$ where $r \in Z_p^*$ we have

$d'_{query1}=(g^{id}Z)^{r_1 + \log_g(Y_1Rh^{s_1})/(x \cdot id)} \cdot (Y_1Rh^{s_1})^{-w/id}=(g^{id}Z)^{r_1} \cdot (Y_1Rh^{s_1})^{\frac{1}{x}}$,

$d_{query1} = (d'_{query1}/g^\theta) \cdot (g^{id}Z)^{r_3} = (g^{id}Z)^{r_4}(Y_1h^{s_0+s_1})^{1/x}$ where $r_4 = r + r_3$.

$d'_{query2}=X^{r_2} \cdot (Y_1Rh^{s_1})^{-/id}=X^{r_1}$

$d_{query2}=d'_{query2} \cdot X^{r_3} = X^{r_4}$.

$d_{query3}=d'_{query3}+s_0=s_0 + s_1$.

It is clear that $d_{queryid,\tau}$ has the same distribution with $d_{id,\tau}$ and the simulator has the ability to construct the private key.

**4.Challenge:** $\beta$ will submit two plaintext $M_0$、$M_1$ with same length which are about to challenge to simulator. The simulator randomly chooses $s \in Z_p^*$, $C'=g^{c^2}$, $C= g^c$, flips a fair binary coin and returns the ciphertext of $M_v (v = 0 \text{ or } v = 1)$ as

$C^*=(\omega^*, C_1^*, C_2^*, C_3^*, C_4^*, C_5^*)=(\omega^*, C'^s, C^s, C'^{sw}, M_vT, C^{s\beta_{k,i}})$

$=(\omega^*, g^{c^2s}, g^{cs}, g^{c^2ws}, M_vT^s, g^{cs\beta_{k,i}})$.

If $\mu = 0, T=e(g,g)^{abc}$. Letting $\gamma=cs$,

$$C^* = (\omega^*, g^{c^2s}, g^{cs}, g^{c^2ws}, M_v e(g,g)^{abcs}, g^{cs\beta_{k,i}}).$$
$$= (\omega^*, g^{c\gamma}, g^{\gamma}, g^{cw\gamma}, M_v e(g,g)^{ab\gamma}, g^{\gamma\beta_{k,i}}).$$

so $C^*$ is the valid ciphertext. Otherwise if $\mu = 1$, $C_5^*$ gives no information about $M_v$.

**5. Phase 2:** The simulator acts exactly as it did in Phase 1.

**6. Guess:** $\beta$ will submit a guess $v'$ of $v$. If $v' = v$, the simulator will output $\mu' = 0$ to indicate that it was given $e(g,g)^{abc}$ to form valid tuple, otherwise it will output $\mu' = 1$ to indicate it was given a random 4-tuple.

The advantage of the adversary in solving *DBDH* problem is $\epsilon/2$. The security proves.

# 5. Analysis

## 5.1 collusion resisting

This section will give the illustration about collusion resisting. Since $E = Me(g, g_4)^{sy_0}$, In order to collusion, they must construct $e(g, g_4)^{y_0}$.

In our system, different user should have different $y_{k,u}$ in the same attribute authority. Assume that there exists a situation that each user's attributes just come from one authority. In our scheme, there is no equality between $y_0$ and $y_{k,u}$. Moreover, different user has different $y_{k,u}$, even each user shares with each other, there is no way to combine them together to compute. Moreover, even attribute authority shares secret values $\{E_{k,i}\}$ and $\{D_{k,i}\}$ where $i \in A_u$ with each other. Since we have $e(g^s, g_2^{(y_0 - \Sigma_{i=1}^K y_{k,u})}) \cdot \prod_{k=1}^K Y_{k,u}^s = e(g^s, g_2^{y_0})$ and

$e(E_{k,i}, D_{k,i}) = e(g, g_2)^{p(0)s} = e(g, g_2)^{y_{k,u}s} = Y_{k,u}^s$. They just have the probability to construct $e(g, g_2)^{y_0}$.

## 5.2 Efficiency analysis

Table 1 gives the contrastive analysis about the efficiency of our multi-authority CP-ABE scheme with other schemes [9,10]. In the table, $k_1$ stands for the number of attributes in $A_c$, $k_2$ stands for the number of attributes in $A_u$. $|G_1|$ stands for the size of the element in $G_1$, $|G_2|$ stands for the size of the element in $G_2$. $K$ is the number of attribute authority in the system. In the cost of encryption and decryption, *Exp* stands for cost of exponentiation and *Pairing* stands for cost of bilinear pairing.

Table 1 COMPARISON WITH OTHER SCHEMES

| Scheme | schemes [9] | schemes [10] | Our scheme |
| --- | --- | --- | --- |
| Private key size | $(2k_2)|G_2|$ | $(k_2+1)|G_1|$ | $|Z_p^*| + (k_2+3)|G_1|$ |
| Ciphertext size | $(2k_1)|G_1| + (1+k_1)|G_2|$ | $3|G_1| + |G_2|$ | $k_1|Z_p^*| + (3+k_1)|G_1| + |G_2|$ |

| Enc time | $(5k_1+2)Exp+k_1Pairing$ | $4Exp$ | $(3+k_1)Exp+Pairing$ |
| --- | --- | --- | --- |
| Dec time | $2k_1Pairing+k_1Exp$ | $4Pairing$ | $4Pairing+(K+2)Exp$ |
| Access structure | LSSS | "And" gate | Access tree |
| Authority | Multi-authority | Multi-authority | Multi-authority |
| Policy hidden | yes | no | yes |
| Collusion resisting | no | no | yes |
| traceable | no | no | yes |

## 6. Conclusion

Our scheme realized Multi-authority ABE encryption.Although the secret values of each authority is leaked in the system ,Collusion is still forbidden. Subsequently, we design a access control scheme with policy hidden,using access tree structure.Once the private key or other secret information leaked, it is traceable who should be responsible for it .Finally, we prove the schemes security with BDBH assumption and analyze its performance with other similar scheme.

## 7. References

[1] Sahai A,Waters B. Fuzzy identity based encryption. In: Proceedings of EUROCRYPT'05. LNCS, 3494. Berlin:Springer, 2005. 457–473

[2] Goyal V, Pandey O, Sahai A, et al. Attribute-based encryption for fine-grained access control of encrypted data. In:Proceedings of the 13th ACM Conference on Computer and Communications Security. New York: ACM Press, 2006.89-98

[3] Bethencourt, J., Sahai, A., Waters, B., 2007. Ciphertext-policy attribute-based encryption. In: IEEE Symposium on Security and Privacy, pp. 321–334.

[4] Chase, M., 2007. Multi-authority attribute based encryption. In: TCC, vol. 4392. LNCS,pp.515–534.

[5] Li J, Huang Q, Chen X, et al. Multi-authority ciphertext-policy attribute-based encryption with accountability. In:Proceedings of the 6th ACM Symposium on Information, Computer and Communications Security, Hong Kong, 2011.386–390

[6] Liu Z, Cao Z, Wong D. White-box traceable ciphertext-policy attribute-based encryption supporting any monotone access structures. IEEE Trans Inf Foren Secur, 2013, 8: 76–88

[7] Zhou J, Cao Z, Dong X, et al. TR-MABE: white-box traceable and revocable multi-authority attribute-based encryp-tion and its applications to multi-level privacy-preserving e-healthcare cloud computing systems. In: Proceedings of the IEEE Conference on Computer Communications, Hong Kong, 2015. 2398–2406

[8] Boneh D, Franklin M. Identity based encryption from the Weil pairing. In: Proceedings of CRYPTO'01. LNCS, 2139.Berlin: Springer, 2001. 213–229

[9] ZHONG Hong,ZHU Wen-long,XU Yan,et al.Multi-authority Attribute-based Encryption Access Control Scheme with Policy Hidden for Cloud Storage[EB/OL]. (2016-09-02)[2017-05-10].https://www.researchgate.net/publication/307627811.

[10] Doshi N,Jinwala D.Constant Ciphertext Length in Multi-authority Ciphertext Policy Attribute based Encryption[C].Proceedings of the 2nd International Conference,Computer and Communication Technology(ICCCT),2011:451-456.